\documentstyle[preprint,aps,epsf]{revtex}
\tighten
\begin{document}
\draft
\title{Heat Kernel Expansions for Distributional Backgrounds}
\author{Ian G. Moss}
\address{
Department of Physics, University of Newcastle Upon Tyne, NE1 7RU U.K.
}
\date{June 2000}
\maketitle
\begin{abstract}
Heat kernel expansion coefficients are calculated for vacuum fluctuations with distributional background potentials and field strengths. Terms up to and including $t^{5/2}$ are presented.
\end{abstract}
\narrowtext
The heat kernel is often used to investigate the effects of vacuum polarisation in quantum filed theory. An important property of the heat kernel is that coefficients in the expansion of the heat kernel in powers of proper time $t$ give information about renormalisation and anomalies. In many cases it is possible to evaluate these coefficients simply from the information provided in the operator, and there is an extensive literature on the subject\cite{dewitt,gilkey}.

Most of the results for the heat kernel coefficients have concentrated on non-singular potentials. However, singular potentials often arise in idealised models. Cosmic strings, for example, have been regarded as distributional line sources in calculations of vacuum energy \cite{dowker}. More recently, in the brane-world models inspired by superstring theory \cite{horova,randall}, the branes are regarded as distributional sources.

Bordag and Vassilevich \cite{bordag} have obtained the most complete results so far on the heat kernel coefficients for delta function potentials, obtaining terms up to $t^{5/2}$. In this letter, similar methods are used to obtain results for gauge fields with distributional field strengths. Such fields appear to be present in the reduction of the Horova-Witten model \cite{horova,lukas}, making the results of topical interest.

The results given here will be for the integrated heat kernel in flat space of $d+1$ dimensions with a delta function potential concentrated on a smoothly embedded surface $\Sigma$ which bounds a region $\Omega$. The heat kernel satisfies
\begin{equation}
\left(\Delta-\partial_t\right)K(x,x',t)=-\delta(t)\delta(x-x')
\end{equation}
where the operator takes the form
\begin{equation}
\Delta=-(\partial+i{\cal A}(x))^2+{\cal V}(x)\label{operator}
\end{equation}
with gauge field ${\cal A}$ and potential ${\cal V}$,
\begin{eqnarray}
{\cal A}=a\theta_\Omega+A\\
{\cal V}=v\delta_\Sigma+V
\end{eqnarray}
The function $\theta_\Omega=1$ for $x\in\Omega$ and zero otherwise. Its derivative is a delta function, $\partial\theta_\Omega=-\delta_\Sigma n$, where $n$ is the unit normal to $\Sigma$. The field strength tensor ${\cal F}_{\mu\nu}={\cal A}_{\mu,\nu}-{\cal A}_{\nu,\mu}$ therefore contains a distributional part,
\begin{equation}
{\cal F}_{\mu\nu}=f_{\mu\nu}\delta_\Sigma+F_{\mu\nu}
\end{equation}
where $f_{\mu\nu}=a_\nu n_\mu-n_\nu a_\mu$.

The integrated heat kernel $K(t)$ is obtained by integrating $K(x,x,t)-K_0(x,x,t)$ over space, where $K_0$ is the free heat kernel. It has an asymptotic expansion of the form
\begin{equation}
K(t)\sim (4\pi t)^{-(d+1)/2}\sum_{n=0}^\infty C_nt^{n/2}\label{pte}
\end{equation}
where, as we shall see, the coefficients can be expressed in terms of integrals of local invariants,
\begin{equation}
C_{n}=\int dx\,b_{n}(x)+\int_\Sigma dx\,c_{n}(x).
\end{equation}

The method for obtaining the heat kernel coefficients is based on perturbation theory \cite{bordag}. Suppose that $\Delta=-\partial^2+{\cal V}(x)$, then
\begin{equation}
K(x,x',t)=K_0(x,x',t)-\int_0^t dt_1\int dx_1K_0(x,x_1,t-t_1){\cal V}(x_1)K(x_1,x',t)
\end{equation}
The iterative solution to this equation is the Born series with terms $K^{(n)}$. For the integrated heat kernel we have,
\begin{eqnarray}
K^{(n)}(t)&=&(-1)^n{t\over n}\int_0^t dt_1\dots\int_0^{t_{n-2}} dt_{n-1}
\int dx_1\dots\int dx_n \nonumber\\
&&{\cal V}(x_1)K_0(x_1,x_2,t-t_1)
\dots {\cal V}(x_n)K_0(x_n,x_1,t_{n-1})\label{born}
\end{eqnarray}
If ${\cal A}=0$, the first term in the series becomes
\begin{equation}
K^{(1)}(t)=-(4\pi t)^{-(d+1)/2}\,t\int dx {\cal V}(x)\label{firstv}
\end{equation}
The second term in the Born series can be simplified by integrating out the intermediate time variable $t_1$,
\begin{equation}
K^{(2)}(t)={t^{1-d}\over2^{d+3}\pi^{d+1/2}}
\int dx dx'e^{-z}U(\case1/2,\case{d+1}/2,z){\cal V}(x){\cal V}(x')\label{secondv}
\end{equation}
where $z=(x-x')^2/t$ and $U(a,b,z)$ is a confluent hypergeometric function.

If the gauge fields are nonzero then it is advantageous to collect together terms which are quadratic in the field strength tensor. Denoting this combination by $K^{FF}$, one finds
\begin{equation}
K^{FF}(t)={t^{1-d}\over2^{d+4}\pi^{d+1/2}}
\int dx dx'e^{-z}U'(\case1/2,\case{d-1}/2,z)
{\cal F}_{\mu\nu}(x){\cal F}^{\mu\nu}(x')\label{secondf}
\end{equation}
where $U'$ is the derivative of $U$ with respect to $z$.

Let us consider the delta function potential $v\delta_\Sigma$ first of all. The second Born approximation reduces to a surface integral
\begin{equation}
K^{(2)}={t^{1-d}\over2^{d+3}\pi^{d+1/2}}
\int_\Sigma dx \int_{\Sigma} dx'e^{-z}U(\case1/2,\case{d+1}/2,z)v(x)v(x')\label{surface}
\end{equation}
For small $t$ the integral is dominated by $x'\approx x$. Let $\xi$ be the unit tangent vector at $x$ to the geodesic joining $x$ to $x'$ in the surface with length $\sigma$. The integrand can be expanded in powers of $\sigma$. Denoting the extrinsic curvature by $k_{ab}$ and the intrinsic ricci curvature by $r_{ab}$, the euclidean distance becomes
\begin{equation}
(x-x')^2=\sigma^2-\case1/{12}(k_{ab}\xi^a\xi^b)^2\sigma^4+\dots
\end{equation}
The surface area element is
\begin{equation}
dx'=(1-\case1/{12}r_{ab}\xi^a\xi^b\sigma^2+\dots)\sigma^{d-1}d\sigma d\xi
\label{area}
\end{equation}
where $\xi$ parameterises a unit sphere in $d$ dimensions. These can be substituted into (\ref{surface}) to obtain an asymptotic expansion in $t$,
\begin{eqnarray}
K^{(2)}&\sim& (4\pi t)^{-(d+1)/2}\int_\Sigma\left\{
{\sqrt{\pi}\over 4}v^2t^{3/2}\right.\nonumber\\
&&\left.+{\sqrt{\pi}\over 32}v\nabla^2vt^{5/2}
-{\sqrt{\pi}\over 128}v^2(k^2-2k_{ab}k^{ab})t^{5/2}+\dots\right\}
\label{sv}
\end{eqnarray}
The operator $\nabla^2$ is the Laplacian on $\Sigma$. The coefficients appearing here, namely $c_{3}$ and $c_{5}$, are in agreement with Bordag and Vassilevich \cite{bordag}.

The same calculation for the gauge fields, using (\ref{secondf}), produces
\begin{eqnarray}
K^{FF}&\sim& (4\pi t)^{-(d+1)/2}\int_\Sigma\left\{
-{\sqrt{\pi}\over 16}f^2t^{3/2}-{1\over 6}f_{\mu\nu}F^{\mu\nu}t^2\right.\nonumber\\
&&\left.-{\sqrt{\pi}\over 256}f\nabla^2ft^{5/2}
+{\sqrt{\pi}\over 1024}f^2(k^2-2k_{ab}k^{ab})t^{5/2}+\dots\right\}
\label{sf}
\end{eqnarray}
where $f^2$ denotes $f_{\mu\nu}f^{\mu\nu}=2a^2-2(a\cdot n)^2$.

Further terms can be evaluated by returning to the Born series (\ref{born}) and taking constant values of the gauge field $a$ and potential $v$. The integrals can be performed for a plane surface $\Sigma$, regarding this as the leading term in the curvature expansion (\ref{area}). The terms in the Born series reduce to expressions of the form
\begin{equation}
K^{(n)}=(4\pi)^{-(d+1)/2}\int_\Omega b^{(n)}+(4\pi)^{-(d+1)/2}\int_\Sigma c^{(n)}
\end{equation}
where, after dropping non-asymptotic terms,
\begin{eqnarray}
c^{(1)}&\sim& -vt\nonumber\\
c^{(2)}&\sim&{\sqrt{\pi}\over 4} v^2t^{3/2}-{\sqrt{\pi}\over 8} a^2t^{3/2}
-{\sqrt{\pi}\over 16} a^4t^{5/2}+{1\over 2}va^2t^2\nonumber\\
c^{(3)}&\sim& -{1\over 6}v^3t^2+{3\sqrt{\pi}\over 16}a^4t^{5/2}
-{1\over 3}va^2t^2-{\sqrt{\pi}\over 8}v^2a^2t^{5/2}\nonumber\\
c^{(4)}&\sim&{\sqrt{\pi}\over 32}v^4t^{5/2}-{29\sqrt{\pi}\over 256}a^4t^{5/2}
+{5\sqrt{\pi}\over 64}v^2a^2t^{5/2}\label{constant}
\end{eqnarray}
The volume integrals depend only on the gauge field $a$ and cancel amongst themselves. Since these terms would violate gauge invariance, the cancellation provides a useful consistency check.

From the preceeding analysis it has emerged that the distributional fields only contribute to surface terms in the heat kernel expansion (\ref{pte}). These surface terms for the operator (\ref{operator}) can now be listed by examination of (\ref{sv}), (\ref{sf}) and (\ref{constant}),
\begin{eqnarray}
c_2&=&-v\\
c_3&=&{\sqrt{\pi}\over 4}v^2-{\sqrt{\pi}\over 16}f^2\\
c_4&=&-{1\over 6}v^3+Vv-{1\over 6}f_{\mu\nu}F^{\mu\nu}+{1\over 12}vf^2\\
c_5&=&{\sqrt{\pi}\over 32}v^4+{\sqrt{\pi}\over 32}v\nabla^2v
-{\sqrt{\pi}\over 128}v^2(k^2-2k_{ab}k^{ab})
-{\sqrt{\pi}\over 4}Vv^2\nonumber\\
&&-{\sqrt{\pi}\over 256}f\nabla^2f
+{\sqrt{\pi}\over 1024}f^2(k^2-2k_{ab}k^{ab})+{3\sqrt{\pi}\over 1024}(f^2)^2
\nonumber\\
&&+{\sqrt{\pi}\over 16}Vf^2-{3\sqrt{\pi}\over 128}v^2f^2
+{\sqrt{\pi}\over 16}vf_{\mu\nu}F^{\mu\nu}
\end{eqnarray}
(Terms involving $V$ have been recovered by multiplying the series by $e^{-Vt}$.)
For comparison, $V$, $v$ and $k_{ab}$ correspond to $-E$, $-V$ and $L_{ab}$ in reference \cite{bordag}. Terms involving $f$ are new, although some of these terms are implicit in other work (e.g. \cite{bordag2}).

The heat kernel coefficients can be used for regularisation in calculations of vacuum fluctuations on distributional backgrounds. If distributional sources are present at a fundamental level, it might also be argued that any divergent terms appearing in the heat kernel expansion must be taken into account in the renormalisation of the theory.

Distributional potentials can also be used to model imperfectly reflecting boundaries, for example as in \cite{frolov}. This means that they are often a more physically realistic choice for modelling the vacuum energy in the Casimir effect than Dirichlet or Robin boundary conditions.

The results can be extended to higher orders in the proper time expansion if that is required. It is also possible to construct alternative expansions of the heat kernel, for example in powers of derivatives as suggested in reference \cite{moss}. Resummation of the potential terms to obtain the first terms in the derivative expansion for $v$ is relatively simple. For constant values of $v$ the results can be checked against the full green function for a delta function potential in one dimension, which is known \cite{blinder}. The corresponding calculation for the gauge fields appears to be more complicated.



\end{document}